# Enhanced Robustness via Loss Engineering in Detuned Non-Hermitian Scattering Systems

Jipeng Xu[1,2*], Yuanhao Mao,[1] Jianfa Zhang,[1,2] Biao Yang[1,2], Ken Liu[1,2*], Zhihong Zhu[1,2*]

[1]College of Advanced Interdisciplinary Studies, National University of Defense Technology, Changsha, Hunan 410073, China.

[2]Nanhu Laser Laboratory, National University of Defense Technology, Changsha 410073, China.

## Abstract:

Non-Hermitian optics has revealed a series of counterintuitive phenomena with profound implications for sensing, lasing, and light manipulation. While the non-Hermiticity of Hamitonians is well-recognized, recent advancements in non-Hermitian physics have broadened to include scattering matrices, uncovering phenomena such as simultaneous lasing and coherent perfect absorption (CPA), reflectionless scattering modes (RSMs), and coherent chaos control. Despite these developments, the investigation has predominantly focused on static and symmetric configurations, leaving the dynamic properties of non-Hermitian scattering in detuned systems largely unexplored. Bridging this gap, we extend certain stationary non-Hermitian scattering phenomena to detuned systems. We delve into the interplay between bi-directional RSMs and RSM exceptional points (EPs), and elucidate the global existence conditions for RSMs under detuning. Moreover, we introduces a novel category of EPs, characterized by the coalescence of transmission peaks, emerging independent with the presence of Hamiltonian EPs. The transmission EPs (TEPs) exhibit flat-top lineshape and can be extended to a square-shaped spectrum when detuning is involved, accompanied by a distinctive phase transition. Significantly, we demonstrate the applications of the TEPs in a one-dimensional coupled cavity system, engineered to enhance sensing robustness against environmental instabilities such as laser frequency drifts, which can exceed 10 MHz. This capability represents a substantial improvement over traditional sensing methods and an important improvement of fragile EP sensors. Our findings not only contribute to the broader understanding of non-Hermitian scattering phenomena but also paves the way for future advancements in non-Hermitian sensing technologies.

# I. INTRODUCTION

Owing to the abundance and controllable non-conservative processes in optical systems, non-Hermitian optics has attracted tremendous interests and inspired a host of exotic phenomenon [1–6] with gain and loss engineering, such as loss-induced transparency [7], laser reviving [8], laser-absorbers [9], unconventional topologies [10,11], to name only a few. A central topic of these advancements is the EP, known as the spectral coalescences of both eigenvalues and eigenvectors of non-Hermitian operators, triggering exciting applications such as improved sensing [12–15], chiral response [16–18] and single mode lasers [19–21].

Recently, the dialogue has expanded beyond classical Hamiltonian EPs to degeneracies associated with the non-Hermitian scattering matrix, particularly absorbing EPs [22,23] that have been observed to support electromagnetic induced transparency lineshapes. Based on these insights, phenomena such as coherent perfect absorption (CPA) and its generalization, RSMs gained their own EPs [24–26], exhibiting a unique quartic lineshape [27] which enables chiral absorption [22] and the suppression of backscattering [26]. However, discussions on these scattering-related EPs have primarily been confined to stationary and geometrically symmetric coupled systems without detuning, thereby omitting the critical dynamic properties essential for applications in sensing and beyond.

EPs are characterized by a nonlinear transition where the parametric response can be unprecedented high [28]. This sensitivity, while beneficial for sensing applications, limits the utility of EP-based sensors due to their susceptibility to fabrication errors and structural instability. Therefore, new strategies for robust EP sensing, such as considering exceptional surfaces [29–31], is quite desired for the practical high sensitive applications.

In this *Letter*, we explore various non-Hermitian scattering phenomenon in symmetric and detuned systems, by both the widely used temporal coupled mode theory (TCMT) and a rigorous scattering matrix approach. We first dissect the stationary behavior of RSMs, encompassing the bi-directional RSMs, RSM EPs, and the global RSM existing conditions with detuning. Next, we propose the concept of transmission exceptional points (TEPs) and its relationship with transmission maximum. We discovered that TEPs persist even in the absence of the Hamitonian EPs, exhibiting distinctive phase transition, detuning symmetry and the square-shaped spectrum. For practical application, we induce a TEP within an one-dimensional non-Hermitian coupled cavity, where the detuning is introduced by linear motions. This configuration yields an abnormal flat-top spectrum, which we propose could significantly enhance the sensing robustness, ensuring operational stability amidst challenges such as probing laser frequency drifts exceeding 10 MHz, which surpasses the capabilities of traditional cavity-based sensors. This work offers a broader understanding of non-Hermitian scattering engineering and paves the way for robust EP sensing, potentially revolutionizing the development of non-Hermitian optical EP devices and their application to other coupled entities.

## II. RESULT

**TCMT model**

Due to the simplicity and comprehensible physical picture, TCMT formalism has been widely used in describing weak-coupled on-chip whispering gallery mode (WGM) cavity systems (Fig. 1a) especially for the modeling of the EPs [4,5]. However, although the spatial one-dimensional cavity system (Fig. 1b) coupled through transmitted light (in contrast to the evanescent coupling of WGMs) would share similar nontrivial behavior with the WGM modes, it still relies more on simple scattering models for analysis, thus lacking descriptions of critical conditions for some special phenomena. We start from the Hamitonian of the coupled mode equations [27]:

$$H = \begin{pmatrix} \omega_1 - i\Gamma_1/2 & \kappa \\ \kappa & \omega_2 - i\Gamma_2/2 \end{pmatrix} \quad (1)$$

Where $\omega_{1,2}$ and $\Gamma_{1,2}$ denote the resonant angular frequencies and total loss rates (including intrinsic losses $\gamma_{1,2}$ and radiative losses $\gamma_{c1,c2}$) of the two sub-cavities.

According to quantum scattering formalism [27,32], the corresponding $2\times 2$ scattering matrix of $H$

within linear approximation could be written as:

$$S(\omega) = \begin{pmatrix} r_1 & t \\ t & r_2 \end{pmatrix} = \begin{pmatrix} 1 - i\dfrac{\gamma_{c1}\Delta_2}{\Delta_1\Delta_2 - \kappa^2} & -i\dfrac{\kappa\sqrt{\gamma_{c1}\gamma_{c2}}}{\Delta_1\Delta_2 - \kappa^2} \\ -i\dfrac{\kappa\sqrt{\gamma_{c1}\gamma_{c2}}}{\Delta_1\Delta_2 - \kappa^2} & 1 - i\dfrac{\gamma_{c2}\Delta_1}{\Delta_1\Delta_2 - \kappa^2} \end{pmatrix} \quad (2)$$

Where $r_{1,2}$ and $t$ denote the reflection and transmission coefficients of two probing ports, respectively. $\Delta_{1,2} = \delta_{1,2} + i\Gamma_{1,2}/2$ with $\delta_{1,2} = \omega - \omega_{1,2}$. The difference between $\omega_1$ and $\omega_2$ is the detuning, which quantifies the extent of resonance mismatch. For instance, in coupled ring microcavities, it can arise from discrepancies in refractive indices or lengths due to temperature variations or stress, or in symmetrically coupled Fabry-Pérot (FP) cavities, from the displacement of the coupling mirrors.

Based on Eqs. (1-2), various types of exceptional points could be predicted when the system keep stationary, i.e., $\delta_1 = \delta_2 = \delta$. For example, the coalescence of eigenvalues of the Hamitonian corresponds to a Hamitonian EP, the degeneracy of the zeros of the scattering matrix correspond to an absorption EP. In particular, CPA EP or RSM EP occurs when the absorption EP or the reflection zeros drop to the real axis of complex plane [22,24,27]. In the following text, we will use loss analysis as a starting point to explore the connections between several non Hermitian scattering phenomena, and then extend them to detuned systems. In order to avoid the errors caused by the TCMT approximation, we adopt the rigorous scattering matrix method to verify the results of the analysis (see supplements).

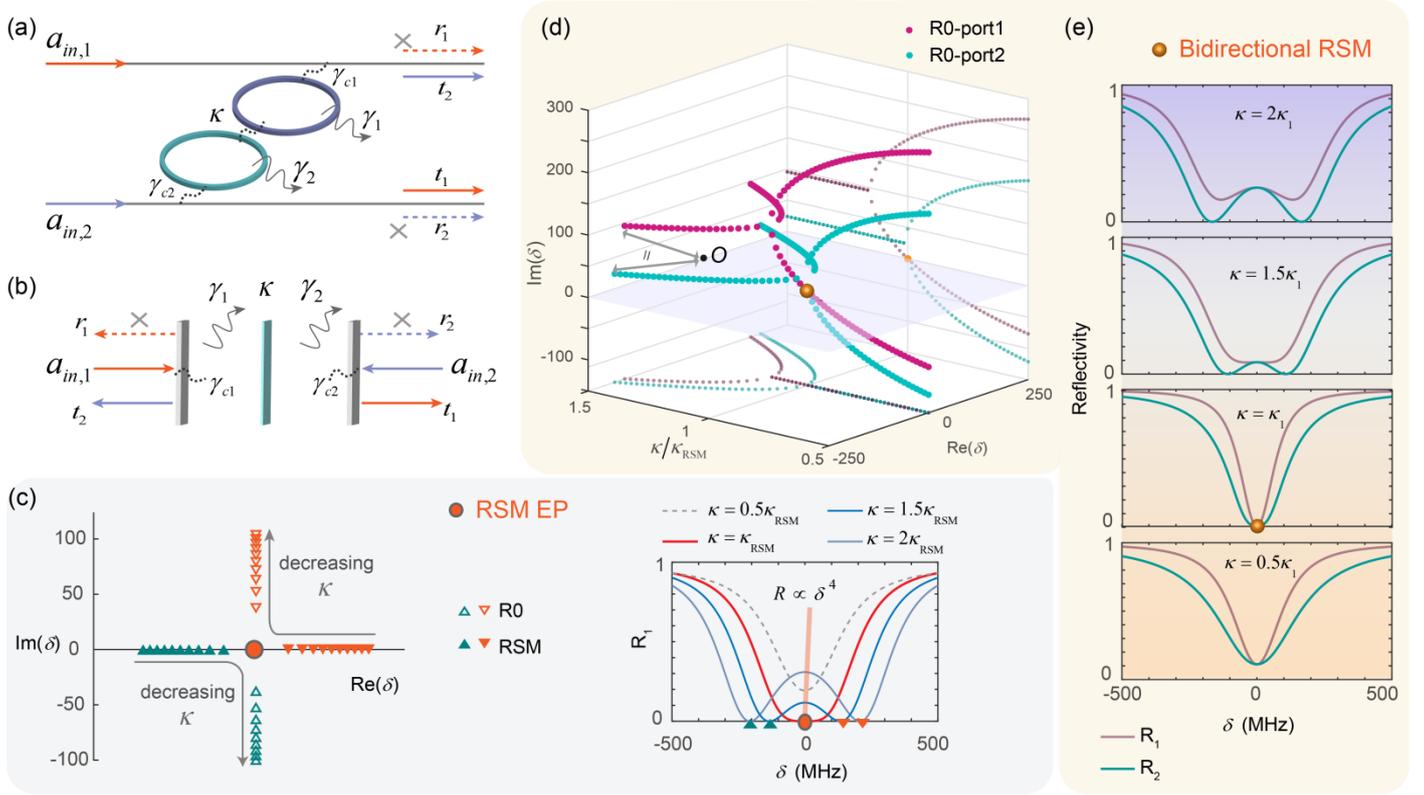

Fig. 1. **RSM EP and bi-directional RSMs in the non-detuned coupled systems.** (a) Coupled WGM microcavities. (b) Coupled FP cavities. (c) With critical coupling condition, an RSM EP occurs at $\kappa = \kappa_1$. Left: In the complex plane, two RSMs approach each other along the real axis as $\kappa$ decreases, subsequently bifurcating into two separate R-zeros. Right: The reflection spectrum as a function of $\kappa$. Parameters: $\gamma_1 = 50$ MHz, $\gamma_2 = 100$ MHz, $\gamma_{c1} = 200$ MHz, $\gamma_{c2} = 350$ MHz. (d) Under the condition of loss proportional, the evolution of real and imaginary parts of the R-zeros with $\kappa$ upon illumination from the two ports. The right and bottom projections respectively show the behavior of quadratic splitting for the real and imaginary parts of R-zeros as $\kappa$ varies. The unique bi-directional RSM occurs at $\kappa = \kappa_1$, highlighted by an orange sphere. The two gray lines on the left side, of equal length, represent equal distances between the R-zeros and the complex plane origin before the R-zeros coalesce. (e) Corresponding reflection spectra of (d) with several specific $\kappa$. It is evident that the proportional losses ensure that the reflections from two ports are always identical for any $\kappa$ when $\delta = 0$. Parameters: $\gamma_1 = 50$ MHz, $\gamma_2 = 100$ MHz, $\gamma_{c1} = 150$ MHz, $\gamma_{c2} = 300$ MHz.

## RSM properties without detuning

Reflection elimination is a desirable target as it could avoid unwanted signal echoes in photonic and microwave networks and enable secure information transmission and analog computation [33,34]. The RSM is used to describe the zero-reflection phenomenon [24], which has been observed in metamaterials [33,35–37], microresonators [22,27,38] and magnonics [34], revealing surprising applications as programmable routers [33], unidirectional invisibility [39], thermal mapping [37] and coherent control of chaos [38,40]. Theoretically, RSMs could be taken as a direct result of critical coupling [41], \emph{e.g.}, the incident coupling equals to the total loss (including inner loss and output loss) for a lossy FP cavity. But the RSMs are usually unidirectional, \emph(i.e.), only one incident direction without reflection, except for purely lossless symmetric FP cavities or central mirror mediated asymmetric cavities [24]. Here, we show that the bi-directional RSMs how to survive in lossy coupled cavities with proper parametric design.

Taking the case of incidence at port 1 as an example, for a lossy cavity without detuning ($\gamma_{1,2,c1,c2} > 0$ and $\delta_1 = \delta_2 = \delta$), $r_{1,2} = 0$ yields:

$$\delta_{R0} = -\frac{1}{4}i(\gamma_1 - \gamma_{c1} + \Gamma_2) \pm \frac{1}{2}\sqrt{-\frac{1}{4}(\gamma_1 - \gamma_{c1} + \Gamma_2)^2 + 4\kappa^2 + (\gamma_1 - \gamma_{c1})\Gamma_2} \qquad (3)$$

RSMs call for purely real $\delta_{R0}$ that could exist in two situations. The first situation is a single RSM at $\delta = 0$, which requires:

$$\kappa_{RSM,1} = \frac{1}{2}\sqrt{(\gamma_{c1} - \gamma_1)\Gamma_2}, \ \gamma_{c1} > \gamma_1 \qquad (4)$$

Similarly we have $\kappa_{RSM,2} = \frac{1}{2}\sqrt{(\gamma_{c2} - \gamma_2)\Gamma_1}$ for the port-2 incidence scenario. The second situation is when $\gamma_{c1} = \gamma_1 + \Gamma_2$ (critical coupling for the port-1 incident), a pair of RSMs located at:

$$\delta = \pm\sqrt{\kappa^2 - \frac{\Gamma_2^2}{4}}, \ \kappa \geq \frac{\Gamma_2}{2} \qquad (5)$$

As shown by the left panel of Fig. 1(c), the R-zeros always hold a purely real number when $\kappa \geq \Gamma_2/2$ if the critical coupling is satisfied. A quartic reflection lineshape (as shown by the right panel of Fig. 1(c)) appears when $\kappa = \Gamma_2/2$ where the two RSMs coalesce at $\delta = 0$ and the so-called RSM EP occurs. The degeneracy behavior of RSM EP are convenient to observe by the coalesce of reflection dips. Once entering the regime of $\kappa < \Gamma_2/2$, all the RSMs vanish.

Moreover, if $r_1 = r_2 = 0$ ($\delta = 0$), we get the bi-directional RSM conditions with $\kappa_{RSM,1} = \kappa_{RSM,2}$:

$$\frac{\gamma_1}{\gamma_2} = \frac{\gamma_{c1}}{\gamma_{c2}} = \eta, \quad \gamma_2, \gamma_{c2} \neq 0 \tag{6}$$

$$\kappa_{\text{bi-RSM}} = \frac{1}{2}\sqrt{\gamma_{c1}\gamma_{c2} - \gamma_1\gamma_2} \tag{7}$$

Where $\eta$ denotes the asymmetry of the system. We term Eq. (6) as a *loss proportional condition* which determines the symmetric reflection for two ports even if the overall structure is asymmetric ($\gamma_1 \neq \gamma_2$, $\gamma_{c1} \neq \gamma_{c2}$). This condition means the product of R-zeros for both incident directions is equal, manifesting as identical distance between R-zeros and the origin before they become purely imaginary (Fig. 1(d)). Decreasing $\kappa$, the RSMs of two ports coincident at the real plane when $\kappa = \kappa_{\text{bi-RSM}}$, where both the ports become reflectionless, as shown by Fig. 1(e).

Above discussions elucidate that the bi-directional RSM EP, defined as the simultaneous occurrence of RSM EPs and bidirectional RSMs, can exclusively manifest in lossless ($\gamma_1 = \gamma_2 = 0$), structural symmetric ($\gamma_{c1} = \gamma_{c1}$) systems. In addition, we note that some previous works have attributed the unidirectional RSMs to the EPs of permuted scattering matrix [35,37,39,42]. Here we can see that the RSMs are always unidirectional except for the Eqs. (6-7) are satisfied. And it has been argued that the EP of a higher dimension permuted scattering matrix is no longer correspond to an RSM [24]。

## Global RSM existence with detuning

Now we turn to the situation of $\delta_1 \neq \delta_2$. For simplicity, we take an anti-symmetric detuning $\omega_2 - \omega_0 = \omega_0 - \omega_1 = \Delta$. The analysis of where the RSM could survive with detuning aids in manipulating the global characteristics of RSMs through detuning. Considering the reflection at port 1 as an example, Fig. 2(a) maps the parameter space of $\kappa - \gamma_{c1}$ to delineate where RSMs occur. In the regime where $\gamma_{c1} < \gamma_1$, the incident cavity is undercoupled, forbidding the existence of RSMs for any values of $\delta$ and $\kappa$. For $\gamma_{c1} > \gamma_1$, RSMs manifest in the blue region where $\kappa \geq \kappa_{RSM}$, but vanish in the orange region where $\kappa < \kappa_{RSM}$.

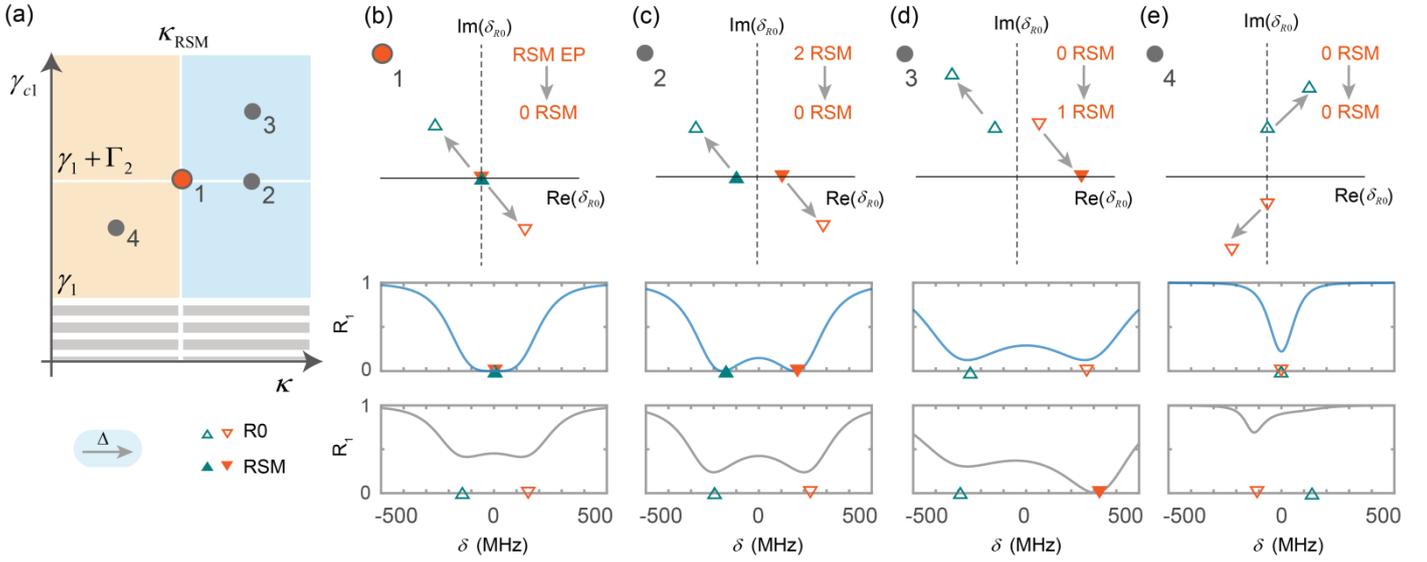

Fig. 2. **RSM existence condition with detuning.** (a) Illustration of $\kappa \sim \gamma_{c1}$ 2-dimensional parameter space. The gray shadowed region forbidden any RSMs, the blue region allows the RSMs beyond $\delta = 0$ and the orange region forbidden them. (b)-(e) Top panels: the motion of R-zeros with detuning for the point 1~4 in (a). Bottom panels: reflection spectra before (blue lines) and after (gray lines) detuning for points 1~4. The gray arrows, hollow triangles, and solid triangles represent detuning, R-zeros, and RSMs, respectively.

Specifically, with $\kappa = \kappa_{RSM}$ and critical coupling (as depicted in Fig. 2(b) at point 1), two R-zeros, initially coalesced at the origin, diverge from the real axis under detuning, leading to the disappearance of RSMs. Similarly, point 2 in Fig. 2(c) represents a scenario where critical coupling is maintained without equality, resulting in the simultaneous vanishing of two previously separated RSMs. At a more general position (point 3), although there was no RSM at the beginning, detuning can lead an R-zero to the real axis, giving rise to an new RSM, as shown in Fig. 2(d). This means that the critical coupling ceases to be a necessary condition for the existence of RSMs at non-zero $\delta$. Finally for point 4, Fig. 2(e) elucidates that two RSMs originally on opposite sides of the real axis cannot be brought to it through detuning, precluding the emergence of new RSMs.

## Transmission maximum and TEP without detuning

The RSMs allow one to design signal transmission without unwanted reflection signals from one or both sides. Furthermore, the high efficiency of the communication devices demands high transmission. We start from the transmission spectrum without detuning. According to Eq. (2) The highest transmission for fixed parameters $\Gamma_{1,2}$ and varied $\kappa$ reads (see supplements for derivation):

$$\max[T(\delta)_{max}] = \frac{\gamma_{c1}\gamma_{c2}}{\Gamma_1\Gamma_2} \tag{8}$$

$$\kappa_{Tmax} = \sqrt{\Gamma_1\Gamma_2}/2, \ \delta=0 \tag{9}$$

Fig. 3(a) gives the comparison of $T(\delta_{max})$ with different $\Gamma_{1,2}$. It is worth noting that when $\Gamma_1 = \Gamma_2$ (here we name it *loss matching condition*), $T(\delta)_{max}$ becomes a constant equals to Eq. (8) that is independent of $\delta$. This means that as long as the losses of the two sub-cavities match each other, the maximum transmission at any frequency position will remain at the same level no matter how the coupling strength changes.

On the other hand, two transmission peaks would coalesce to a single at $\delta = 0$ when (see supplements for derivation):

$$\kappa_{TEP} = \frac{\sqrt{\Gamma_1^2 + \Gamma_2^2}}{2\sqrt{2}} \tag{10}$$

Eq. (10) defines a transmission EP (TEP) characterized by the square-root dependence of the transmission peak frequency on the coupling strength by, as shown by the top panel of Fig. 3(b) and 3(c). To our best knowledge, this kind of EP has not been well-described before. In contrast to the description in ref. [27], the transmission peak is **not** actually situated at the poles of the S-matrix, and the point where the transmission peaks merge is not the coalesce of the poles. An intuitive counterexample emerges when $\Gamma_1 = \Gamma_2$, under which, for a finite $\kappa$, scattering poles never coalesce, hence precluding the presence of Hamitonian EPs. However, as demonstrated in Fig. 3(c), TEPs do exist and merge with $T_{max}$ as Eq. (9) is equivalent to Eq. (10).

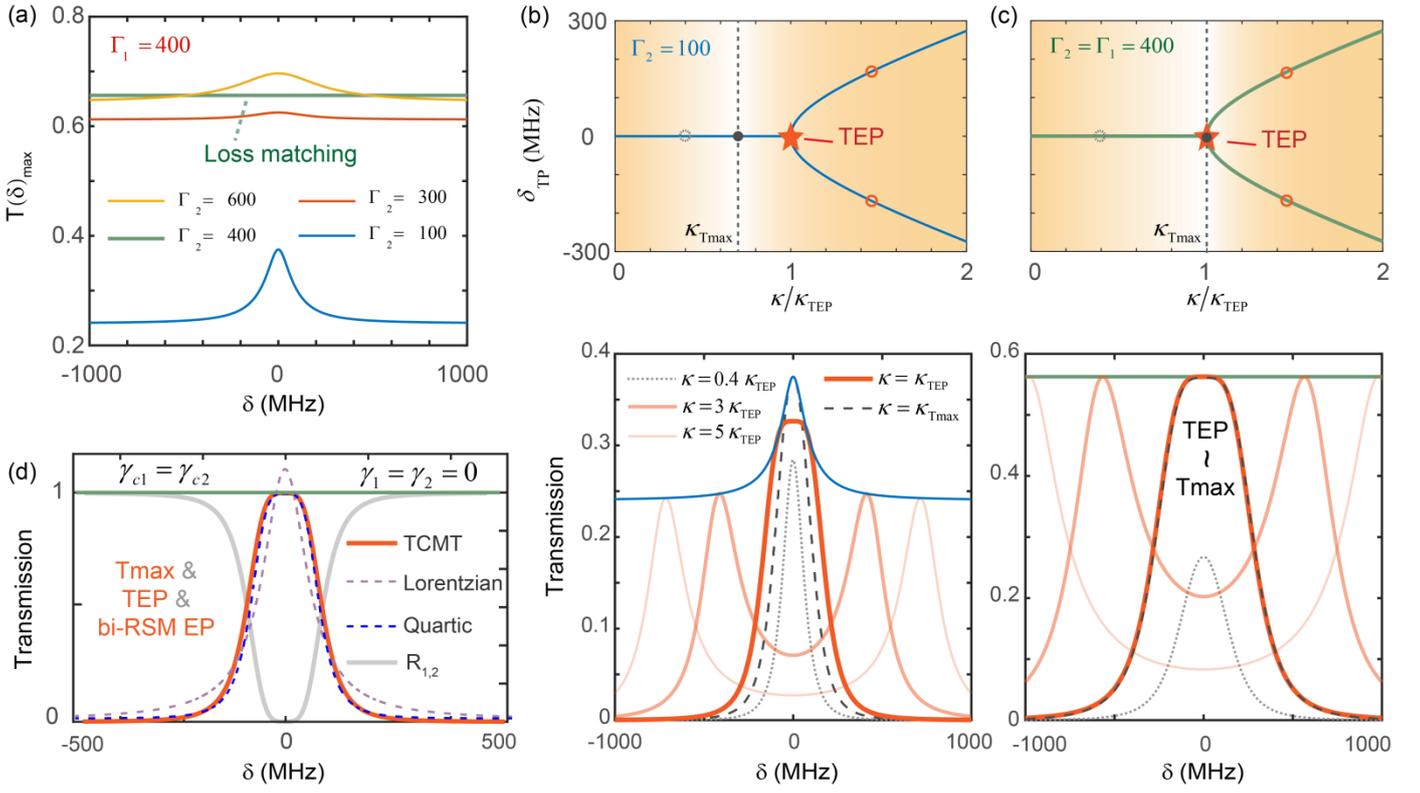

Fig. 3. **Transmission maximum and TEPs without detuning.** (a) Dependence of $T_{max}$ on $\delta$ with different $\Gamma_2$. The green line shows that when loss match, the maximum transmission is a constant. (b) Top panel: Transmission peak frequencies as a function of coupling strength when $\Gamma_1 \neq \Gamma_2$. The TEP occurs when $\kappa = \kappa_{TEP}$. Bottom panel: The transmission spectrum with different $\kappa$ and they coalesce at a TEP, featuring by a quartic lineshape. As $\kappa$ continues to increase, the single peak transmission touches the maximum at $\kappa_{Tmax}$ and then gradually decrease. The vertical gray dotted lines denote the coupling strength corresponds to the maximum transmission. (c) When loss matching condition $\Gamma_1 = \Gamma_2$ holds, $\kappa_{TEP} = \kappa_{Tmax}$, where the top of the quartic line shape coincides with $T_{max}$. (d) A TEP meets a bi-directional RSM EP when $\Gamma_1 = \Gamma_2$ and the inner loss is totally removed. The fitting results use quartic (blue dashed) and Lorentzian (gray dashed) functions. Parameters: $\gamma_1 = 100$ MHz, $\gamma_2 = 50$ MHz, $\gamma_{c1} = 300$ MHz, $\gamma_{c2}$ is the variable.

The transmission spectra with different $\kappa$ (bottom panels of Fig. 3(b-c)) demonstrate that the prominent feature of TEPs is the flat-top quartic lineshape ($T(\delta=0) - T(\delta) \propto \delta^4$), which is similar to the CPA EP [27] and RSM EP [24] but not rely on special loss settings. When $\Gamma_2 = 100$ MHz, as $\kappa$ gradually decreases from $5\kappa_{TEP}$, the two transmission peaks ($\delta_{TP}$) gradually approach and increase. Once the system bypasses the $\kappa_{TEP}$, they will merge into a single peak and continue to increase until $\kappa_{Tmax}$, and then gradually decrease with larger $\kappa$. Especially, when $\Gamma_1 = \Gamma_2$, the quartic TEP directly corresponds to the maximum of transmission spectrum.

According to Eq.(8), only if $\gamma_1 = \gamma_2 = 0$, emph{i.e.}, the cavity is totally lossless, $\max[T(\delta)_{\max}] = 1$ with $\kappa = \sqrt{\gamma_{c1}\gamma_{c2}}/2$, at which the $T_{\max}$ is coincident with the bi-directional RSM. Furthermore, if the coupling losses are symmetric ($\gamma_{c1} = \gamma_{c2}$), the critical coupling, proportional losses and loss matching condition coincident and thus the bi-directional RSM EP, $T_{\max}$ and TEP occurs simultaneously (Fig. 3(d)). Beyond this critical scenario, the one-port RSM and maximum transmission could also coexist when the incident cavity is lossless ($\gamma_1$ or $\gamma_2$ vanishes), which is of great significance for non-Hermitian filter design.

In practical optical applications, people are more interested in the spectrum characteristics rather than the commonly discussed resonant modes. In systems with inexact parity-time symmetry, the scattering poles and zeros do not form complex conjugate pairs [43], leading to an inherent deviation of the spectrum peaks from the resonant modes in purely lossy systems. This deviation has recently inspired the idea of using the spectral degeneracy point to enhance the signal-to-noise ratio of non-Hermitian sensors [44]. In this binary system, the Hamitonian EP condition reads $\kappa_{\text{HEP}} = |\Gamma_1 - \Gamma_2|/4$ [5], which is always smaller than $\kappa_{\text{TEP}}$ except for the exact parity-time symmetric scenario ($\Gamma_1 = -\Gamma_2$) where $\kappa_{\text{TEP}} = \kappa_{\text{HEP}}$. For example, consider a specific passive non-Hermitian coupled cavity only with unilateral loss, it can be derived that $\kappa_{\text{TEP}} = \sqrt{2}\kappa_{\text{HEP}}$. Especially in the case of $\Gamma_1 = \Gamma_2$, the Hamitonian EP disappears, whereas the TEP still exists and maintain the square-root properties. Furthermore, due to its merging with $T_{\max}$, the resulting flat transmission peaks (varying with $\kappa$) facilitates calibration and control during practical measurements.

## Square-shaped scattering spectrum with detuning

Next, we demonstrate that the quartic transmission spectrum characterized by TEP could be generalized to a square-shaped transmission map and the square-root dependence of $\delta_{TP}$ on $\kappa$ is independent with its relationship with the detuning. Considering the anti-symmetric detuning situation where $\delta_1 = \delta - \Delta$ and $\delta_2 = \delta + \Delta$ with $\Gamma_1 = \Gamma_2 = \Gamma$, the transmission spectrum given by TCMT could be simplified as:

$$T(\delta, \Delta) = \frac{\kappa^2 \gamma_{c1} \gamma_{c2}}{C(\delta, \Delta)}$$

$$C(\delta, \Delta) = [\delta^2 - \Delta^2 + (\frac{\Gamma^2}{4} - \kappa^2)]^2 + \Gamma^2(\kappa^2 + \Delta^2) \tag{11}$$

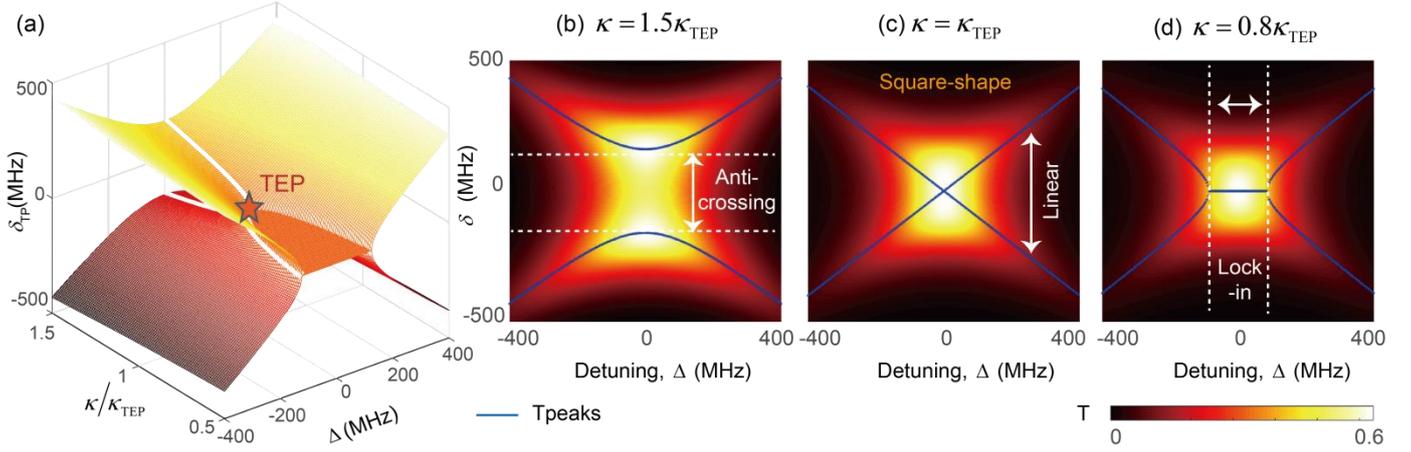

Fig. 4. **Transmission detuning symmetry and the TEP phase transition.** (a) Dependence of the transmission peaks on detuning and mutual coupling strength. (b)-(d) Transmission spectra versus detuning and incident frequency with (a) $\kappa > \kappa_{TEP}$, (b) $\kappa = \kappa_{TEP}$, (c) $\kappa < \kappa_{TEP}$. The blue lines denote the transmission peaks. As $\kappa$ decreases, the quadratic dependence of transmission peak frequency on the detuning transform into linear and then square-root dependence.

It is obvious that $T(\delta, \Delta) = T(-\delta, \Delta)$ which indicates the transmission is always symmetric for any $\Delta$. It should be emphasized that the detuning symmetry for $T$ requires both the loss matching condition and anti-symmetric detuning which is natural in the one-dimensional coupled cavity (Without this premise the special symmetry would break, see supplements). As the parameters in Eq. (11) are all positive, the transmission peak points reads:

$$\delta_{TP} = \pm \sqrt{\Delta^2 + (\kappa^2 - \frac{\Gamma^2}{4})} \tag{12}$$

The discrepancy between transmission peaks and S-matrix poles could be elucidated here: the transmission peaks only demand that $C$ attains a local minimum, while the scattering poles necessitate that $C(\delta, \Delta) = 0$. Based on Eq. (12), we give a three-dimensional dependency amongst $\delta_{TP}$, $\Delta$ and $\kappa$, as depicted in Fig. 4(a). A phase transition phenomenon emerges around the critical condition distinguishing $T_{\max}$ and TEP, $\kappa = \Gamma/2$, at which the transmission peak points vary linearly with detuning. This distinction effectively bifurcaites the behavior into regimes characterized by quadratic ($\kappa > \Gamma/2$) and square-root ($\kappa < \Gamma/2$) dependencies, as illustrated by Fig. 4 (b-d). The TEP demonstrates a detuning response markedly distinct from that observed in Hamitonian EPs [5,14,45]. For both phenomena, as $\kappa$ increases, the frequency of the transmission peak and the eigenfrequency undergo quadratic changes with detuning, exhibiting an anti-crossing in the transmission maps. Diminishing $\kappa$, in the former case, leads to a linear variation of the transmission peak frequency with detuning at the TEP, followed by the emergence of a lock-in-like region [46] insensitive to detuning upon further reduction. Conversely, in the latter case, the eigenfrequency responds to detuning with a square-root behavior at the Hamitonian EP, evolving towards a linear response as $\kappa$ continues to decrease. Fig. 4(d) reveals that outside the detuning-insensitive zone associated with TEP, an enhanced square-root response is observed, akin to the EP-enhanced Sagnac effect [14,15]. Moreover, this detuning-insensitive zone has a boundary as the coupling decreases, which, according to Eq. (12), occurs at $\Delta = \Gamma/2$.

It is noteworthy that the detuning symmetry elucidated here and the square-shaped transmission spectra depicted in Fig. 4(c) are also present in the reflection spectra. However, their occurrence in the reflection spectra necessitates an additional condition of critical coupling (See supplements). Additionally, the phase transition of TEP with loss matching condition, while superficially similar to the band characteristics of non-Hermitian photonic crystal slabs [47], is fundamentally different. There the horizontal axis represented by the wave vector essentially reflects the coupling strength but here the horizontal axis of the TEP spectrum map is detuning.

# Roubustness given by the square-shaped spectrum

Metrological applications utilizing a single cavity critically depend on a movable mirror to transduce displacement and vibration into detectable variations in the light wave phase or intensity [48]. Nonetheless, the constancy of the probe laser frequency, indispensable for high-precision measurements, is compromised by mechanical instabilities and thermal perturbations. To counteract these fluctuations, sophisticated frequency stabilization mechanisms, including injection locking and active feedback loops [49], are mandated. However, these supplementary apparatus often contribute to system bulkiness, rendering the minimization of their footprint a paramount objective in the design of streamlined metrological devices.

The detuning of the one-dimensional cavity consists of two sub-cavities comes from the asymmetric cavity lengths, which could also be used for measuring displacements or vibrations by the central mirror movements. For example, it is an effective platform to achieve quantum Fock state read-out [50]. Here we take the simplest case of $\gamma_1 = \gamma_2 = 0$ and $\gamma_{c1} = \gamma_{c2}$, where both the R and T spectrum have quartic properties to expound our robust metrology protocol.

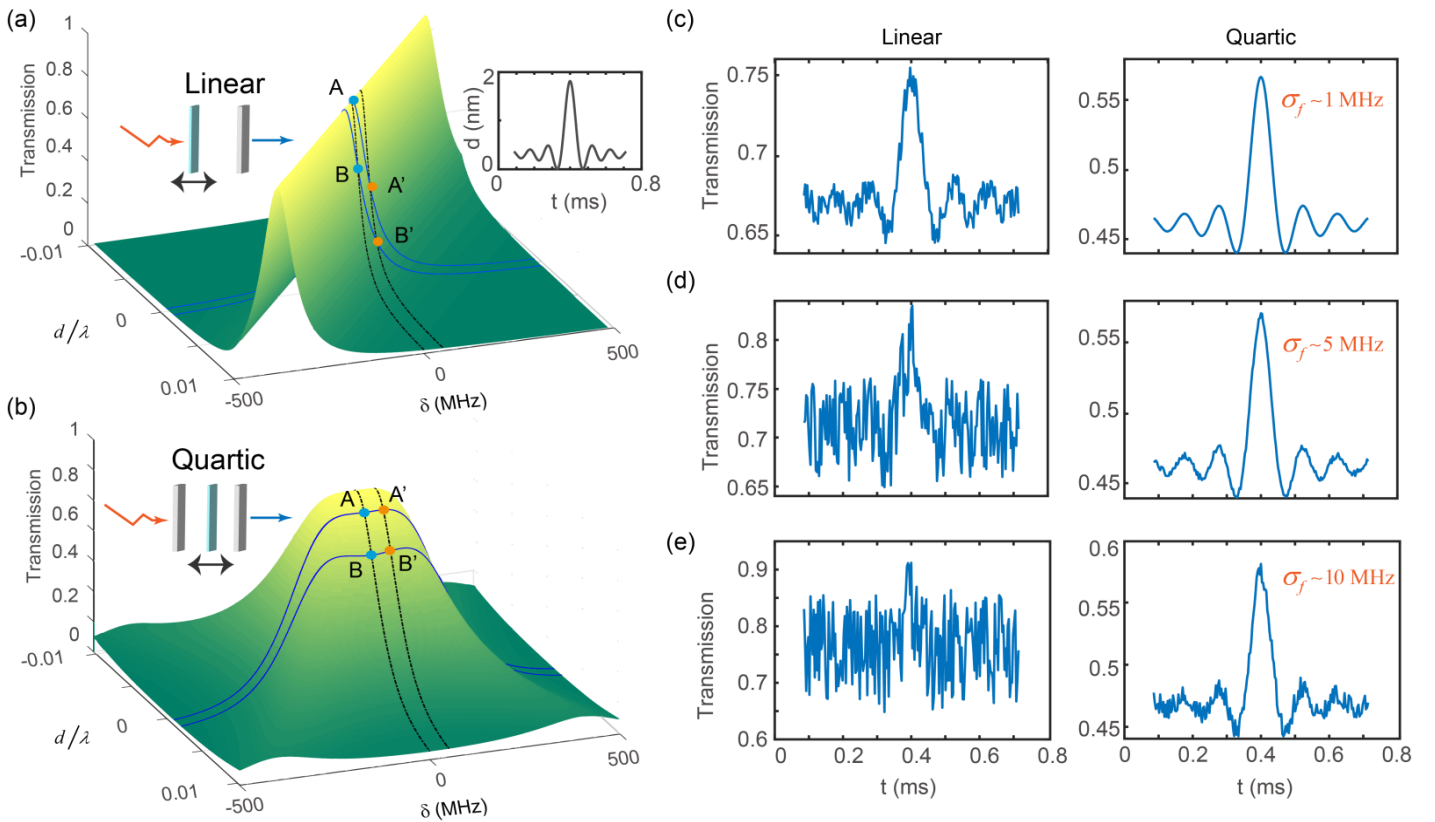

FIG. 5. **Comparison of displacement sensing via linear and square-shaped spectra.** The three-dimensional plots illustrate the dependence of transmission on displacement and incident frequency for (a) a single FP cavity and (b) a non-Hermitian coupled cavity, respectively. Left insets depict the corresponding sensing systems. Right inset of (a) gives the displacement signal. In (a) and (b), points A and B represent the light intensities at two displacement positions under unperturbed frequency conditions, while A' and B' denote the responses at an alternate frequency position following perturbation. Panels (c)-(e) display the detuning responses of both systems to an input displacement signal, subject to random frequency disturbances of magnitudes (c) 1 MHz, (d) 5 MHz, and (e) 10 MHz, respectively.

Fig. 5(a) illustrates the relationship among the transmission of a single FP cavity, incident frequency and central mirror displacement, widely applied in fundamental detection devices such as displacement and acceleration detection, due to its high sensitivity and excellent linear response Typically, the detection approach involves irradiating the cavity with a narrow-band laser near one peak, such as at point A (not necessarily at the peak), whereupon displacement causes a reduction in light intensity to point B. However, should the laser frequency jitter, the light intensity oscillates between points A, B, A', and B', rendering the displacement readout unstable. In contrast, employing the proposed square-shaped spectrum for detection, as depicted in Fig. 5(b), ensures stable light intensity response to displacement within a certain range of frequency jitter, as the intensity at points A and B is virtually indistinguishable from that at A' and B'. For instance, applying a sinusoidal vibration signal $d(t) \propto \text{sinc}(t)$ to the central mirror, both detection methods can replicate it through variations in light intensity $T(t)$, yet exhibit marked differences under varying degrees of frequency disturbance, as shown in Figs. 5(c)-(e) for $\sigma_f$ equals to 1 MHz, 5 MHz, and 10 MHz. It is evident that with increasing frequency perturbation, the response of the Hermitian cavity quickly becomes overwhelmed by noise, whereas the non-Hermitian cavity demonstrates robust resistance to frequency noise. These results highlight the self-stabilizing capability of square spectral response-based displacement or vibration detection, potentially reducing the footprint required for frequency stabilization equipment.

# Conclusion

In conclusion, this study broadens the scope of non-Hermitian scattering explorations to encompass dynamic and detuned systems. The robustness of our findings is demonstrated through both analytical methods, employing TCMT and a rigorous scattering matrix approach. Leveraging three specific loss conditions, we explored several unique non-Hermitian scattering phenomena and the connections among them, both static and dynamic. The critical coupling condition ensures that R-zeros remain on the real axis in the complex plane until their eventual coalescence. The loss proportional condition mandates identical reflection at the central frequency for two ports, highlighting the intricacies of system symmetry. Most notably, the loss matching condition unveils multifaceted functionalities: without detuning, it aligns the TEPs with the transmission maxima, maintaining constant maximum transmission across the spectral range. With detuning, this condition preserves the transmission detuning symmetry, showcasing its versatility.

A groundbreaking discovery in our study is the ubiquitous quartic spectral lineshapes, which previously identified as the unique behavior of RSM EPs or CPA EPs. In the detuned case, it evolves to a square-shaped spectral map and exhibit distinct phase transition properties. We designed a practical application scenario that significantly enhances the robustness of displacment sensing against environmental instabilities, such as laser frequency drifts. This represents a considerable improvement over existing sensing technologies, reinforcing the practicality and utility of EP-based sensors. Overall, this work not only expands the theoretical framework of non-Hermitian scattering but also lays the groundwork for the development of more sophisticated and dependable non-Hermitian sensors, paving the way for future innovations in optical physics.

# Supplementary material of Enhanced Robustness via Loss Engineering in Detuned Non-Hermitian Scattering Systems

## S1. Rigorous scattering analysis for the one-dimensional coupled cavity

In the one-dimensional scattering system, the phase accumulation during a round trip in the left or right sub-cavity are given by:

$$\varphi_L(v, \Delta x) = \exp[-2ik(l + \Delta x)]$$
$$\varphi_R(v, \Delta x) = \exp[-2i(k - i\gamma)(l - \Delta x)]$$
(S1)

Calculate the effective reflection and transmission coefficients of the middle and right mirrors as a whole:

$$r_{eff} = r_c + \frac{r_2 t_c^2 \varphi_R}{1 - r_2 r_c \varphi_R}$$

$$t_{eff} = \frac{t_2 t_c \varphi_R}{1 - r_2 r_c \varphi_R}$$
(S2)

Then the total reflection and transmission spectra under left incident waves take the form:

$$r_L = r_1 + \frac{r_{eff} t_1^2 \varphi_L}{1 - r_1 r_{eff} \varphi_L}, \quad R_L = |r_L|^2$$

$$t_L = \frac{t_{eff} t_1 \varphi_L}{1 - r_1 r_{eff} \varphi_L}, \quad T_L = T_R = |t_L|^2$$
(S3)

The rigorous SMA could help us check the efficiency of TCMT for the one-dimensional cavity. The radiative coupling through the side mirrors with reflection coefficients $r_L$ and $r_R$ are $\gamma_{c1,c2} = -2\ln(r_{L,R})/cL$, where $L$ is the whole cavity length. It should be noted that the mutual coupling constant here is derived as $\kappa = \mathrm{acosh}(1/r_c)/cL$ based on the Hamitonian EP condition with the help of the scattering matrix approach. As shown by Fig. S1, the TCMT works well at weak coupling regime before $\kappa_{1,2}$, but when the coupling becomes stronger, a little discrepancy occurs.

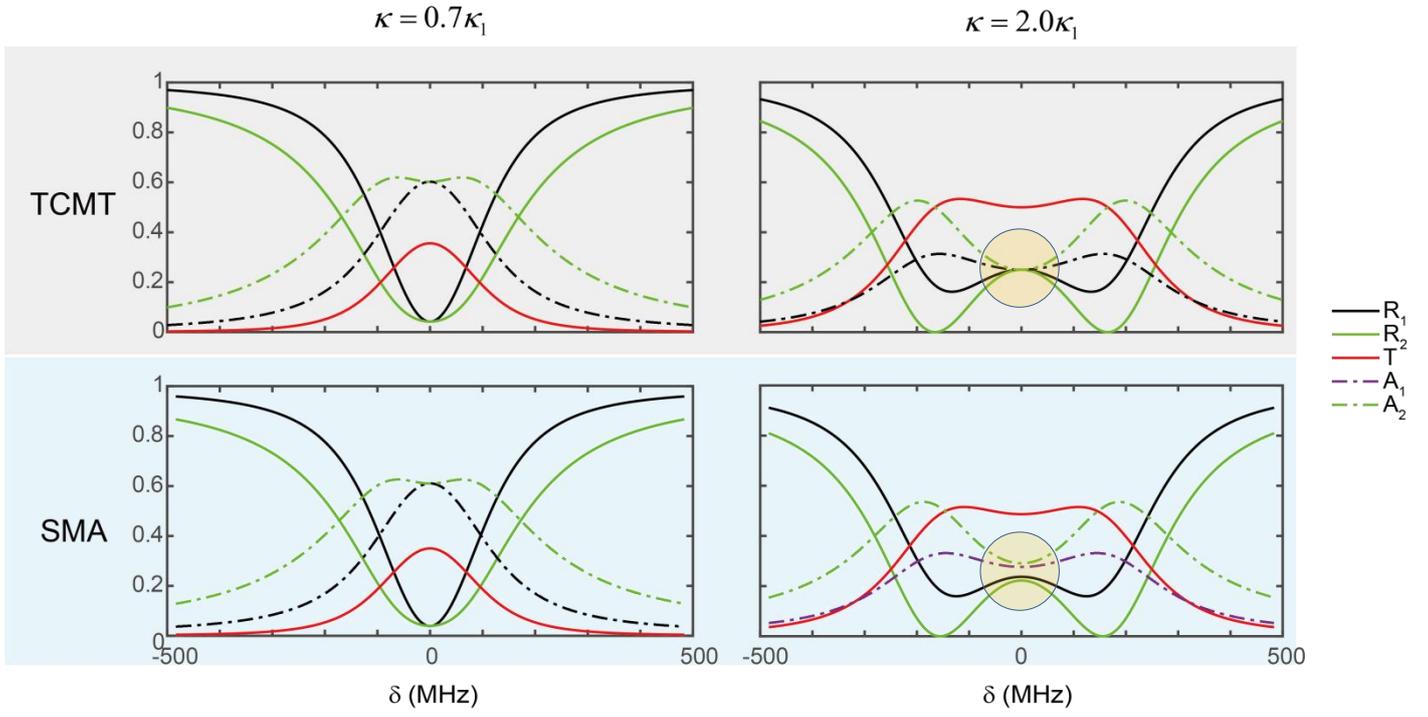

Fig. S1. Comparison of reflection, transmission and absorption spectra of the coupled FP system obtained by TCMT and SMA under weak and strong coupling regimes. Beyond $\kappa > \kappa_1$, a deviation (shown by the light yellow region) between the two arises, indicating that the TCMT approximation gradually fails. Parameters: $\gamma_1 = 50$ MHz, $\gamma_2 = 100$ MHz, $\gamma_{c1} = 150$ MHz, $\gamma_{c2} = 300$ MHz.

## S2. Derivation of transmission maximum and the transmission EP

**Transmission maximum.** Considering the two sub-cavities do not have detuning and the central mirror is transmissive ($\kappa \neq 0$), the transmission spectrum given by TCMT is written as:

$$T(\delta) = \frac{\kappa^2 \gamma_{c1} \gamma_{c2}}{\left[(\delta + i\frac{\Gamma_1}{2})(\delta + i\frac{\Gamma_2}{2}) - \kappa^2\right]\left[(\delta - i\frac{\Gamma_1}{2})(\delta - i\frac{\Gamma_2}{2}) - \kappa^2\right]} = \frac{\gamma_{c1}\gamma_{c2}}{\kappa^2 + \frac{B(\delta)}{\kappa^2} - 2\delta^2 + \frac{\Gamma_1 \Gamma_2}{2}} \quad (S4)$$

where $\delta = \omega - \omega_0$, $\Gamma_1 = \gamma_1 + \gamma_{c1}$, $\Gamma_2 = \gamma_2 + \gamma_{c2}$, $B(\delta) = (\delta^2 + \frac{\Gamma_1^2}{4})(\delta^2 + \frac{\Gamma_2^2}{4})$. for any $\kappa$ and specific $\delta$, the maximum transmission could be obtained:

$$T(\delta) \leq T(\delta)_{max} = \frac{\gamma_{c1}\gamma_{c2}}{2(\sqrt{B(\delta)} - \delta^2) + \frac{\Gamma_1\Gamma_2}{2}} \quad (S5)$$

Where $\kappa^2 = \sqrt{B}$ corresponds to the maximum. Eq. (S5) gives Fig. 2(a) in the main text. It is obvious that $T(\delta = 0)_{max} \geq T(\delta)_{max}$ and the equal sign holds only if the loss matching condition $\Gamma_1 = \Gamma_2$ is satisfied, where $T(\delta = 0)_{max} = \frac{\gamma_{c1}\gamma_{c2}}{\Gamma_1 \Gamma_2}$ with $\kappa_{Tmax} = \frac{\sqrt{\Gamma_1 \Gamma_2}}{2}$, as depicted by the green line in Fig. 2(a) in the main text. For the symmetric case with $\Gamma_1 = \Gamma_2 = \Gamma$, $\gamma_1 = \gamma_2 = \gamma$, $\gamma_{c1} = \gamma_{c2} = \gamma_c$, the device could achieve a highest transmission $T(\delta = 0)_{max} = (\gamma_c/\Gamma)^2$ when $\kappa = \Gamma/2$.

**Transmission EP.** For any $\delta$ and specific $\kappa$, according to Eq. (S4), the peak frequency of transmission could be obtained by find the minimum of $M = \kappa^4 + B - 2\delta^2 \kappa^2$. It can be simplified as:

$$\begin{aligned} M &= \delta^4 + (\frac{\Gamma_1^2 + \Gamma_2^2}{4} - 2\kappa^2)\delta^2 + \kappa^4 + \frac{\Gamma_1^2 \Gamma_2^2}{16} \\ &= [\delta^2 - (\kappa^2 - \frac{\Gamma_1^2 + \Gamma_2^2}{8})]^2 + \kappa^4 + \frac{\Gamma_1^2 \Gamma_2^2}{16} - (\kappa^2 - \frac{\Gamma_1^2 + \Gamma_2^2}{8})^2 \end{aligned} \quad (S6)$$

The peak frequency reads:

$$\delta_{peak} = \pm\sqrt{\kappa^2 - \frac{\Gamma_1^2 + \Gamma_2^2}{8}} \quad (S7)$$

It follows that the two transmission peaks coalesce at:

$$\kappa_{\text{TEP}} = \frac{\sqrt{\Gamma_1^2 + \Gamma_2^2}}{2\sqrt{2}} \tag{S8}$$

When $\kappa < \kappa_{\text{TEP}}$, the single peak remains at $\delta = 0$. The prominent feature of TEP is the quartic transmission spectrum, which can be proved by:

$$T(\delta = 0) - T(\delta)$$
$$= \frac{\kappa^2 \gamma_{c1} \gamma_{c2} \delta^4}{(\kappa^4 + \frac{\Gamma_1 \Gamma_2}{2} \kappa^2 + \frac{\Gamma_1^2 \Gamma_2^2}{16})(\delta^4 + \kappa^4 + \frac{\Gamma_1 \Gamma_2}{2} \kappa^2 + \frac{\Gamma_1^2 \Gamma_2^2}{16})} \propto \delta^4 \tag{S9}$$

## S3. Detuning symmetry for transmission spectra

The detuning symmetry of transmission spectrum have two premises:

1) loss matching condition $\Gamma_1 = \Gamma_2 = \Gamma$;

2) anti-symmetric detuning $D_1 = -D_2 = \Delta$. ($D_{1,2} = \delta_{1,2} - \delta$).

Fig. S2 plots the comparison of the symmetric situation and two asymmetric situations without the premise 1) or 2).

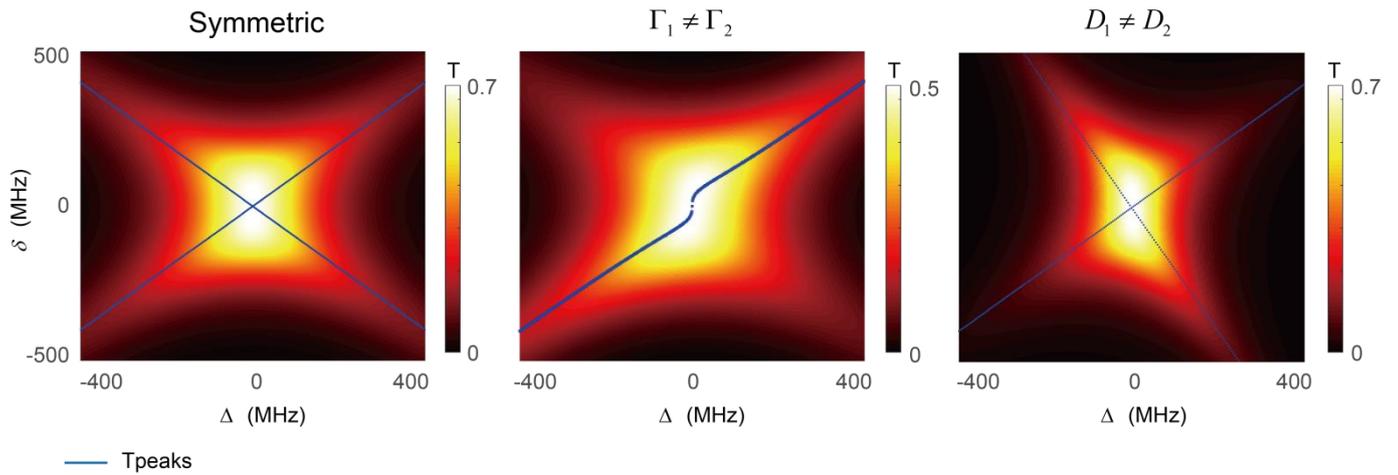

Fig. S2. The transmission map as a function of $\Delta$ and $\delta$. (a) The symmetric case. (b) Non-symmetric case with $\Gamma_1 = \frac{3}{4}\Gamma_2$. (c) Non-symmetric case with $D_2 = 2D_1$.

## S4. Detuning symmetry for reflection spectra

The reflection spectrum map with detuning could also be symmetric. We derive its criteria by solving $R_1(\delta, \Delta) = R_1(-\delta, \Delta)$ ($\Delta \neq 0$), which yields (we have $\Delta_1(\delta, \Delta) + \Delta_1^*(-\delta, \Delta) = \Delta_1^*(\delta, \Delta) + \Delta_1(-\delta, \Delta) = -2\Delta$, $\Delta_2(\delta, \Delta) + \Delta_2^*(-\delta, \Delta) = \Delta_2^*(\delta, \Delta) + \Delta_2(-\delta, \Delta) = 2\Delta$):

$$(1-i\frac{\gamma_{c1}\Delta_2}{\Delta_1\Delta_2 - \kappa^2})(1+i\frac{\gamma_{c1}\Delta_2^*}{\Delta_1^*\Delta_2^* - \kappa^2}) = (1-i\frac{\gamma_{c1}(\Delta_2 - 2\Delta)}{(-2\Delta - \Delta_1)(2\Delta - \Delta_2) - \kappa^2})(1+i\frac{\gamma_{c1}(\Delta_2^* - 2\Delta)}{(2\Delta - \Delta_1^*)(-2\Delta - \Delta_2^*) - \kappa^2})) \quad (S9)$$

It follows that:

$$\begin{aligned}\Gamma_1 &= \Gamma_2 \\ \gamma_1 &= 0\end{aligned} \quad (S10)$$

In order for symmetric reflection maps to be realized, both of the loss matching condition and critical coupling condition must be simultaneously satisfied. An intuitive understanding is that when critical coupling is satisfied, the critical coupling makes the distribution of the R-zeros being symmetrically centered around the origin. Conversely, symmetric transmission maps ($T(\delta, \Delta) = T(-\delta, \Delta)$) necessitate solely the satisfaction of $\Gamma_1 = \Gamma_2$, as the numerator of $T$ is independent of $\delta$ and $\Delta$.

It should be emphasized that the R-zeros could not intuitively correspond to the reflection behavior before it comes to the real axis. As a result, As evidence, considering the symmetric detuning $\Delta$ caused by the movement of central mirror. $r_1 = 0$ yields that:

$$(\delta - \Delta)(\delta + \Delta) + i\frac{\Gamma_2}{2}(\delta - \Delta) + i(\frac{\Gamma_1}{2} - \gamma_{c1})(\delta + \Delta) - \frac{1}{4}\Gamma_1\Gamma_2 - \kappa^2 + \frac{\Gamma_2}{2}\gamma_{c1} = 0 \quad (S11)$$

It follows that:

$$\delta_{\text{R-zero}} = -\frac{1}{4}i(\gamma_1 - \gamma_{c1} + \Gamma_2) \pm \frac{1}{2}\sqrt{-\frac{1}{4}(\gamma_1 - \gamma_{c1} + \Gamma_2)^2 + \Gamma_1\Gamma_2 + 4\kappa^2 - 2\Gamma_2\gamma_{c1} + 4\Delta^2 - 2i(\gamma_1 - \gamma_{c1} - \Gamma_2)\Delta}$$

(S12)

In Fig. S3, we present the $R_1$ spectrum for with/without detuning when $\gamma_1 = \gamma_{c1} + \Gamma_2$ where Eq. (S12) is simplifies as $\delta_{\text{R-zero}} = -\frac{1}{2}i\Gamma_2 \pm \sqrt{\kappa^2 + \Delta^2}$. In such scenarios, detuning merely results in the lateral, opposed displacement of the two R-zeros in the complex plane, without introducing variation in the imaginary components. Nonetheless, the two reflection peaks forfeit their detuning symmetry when $\Delta$ comes, indicating that the magnitude of the imaginary parts of the zeros does not correspond to the reflectance peak heights or bandwidths.

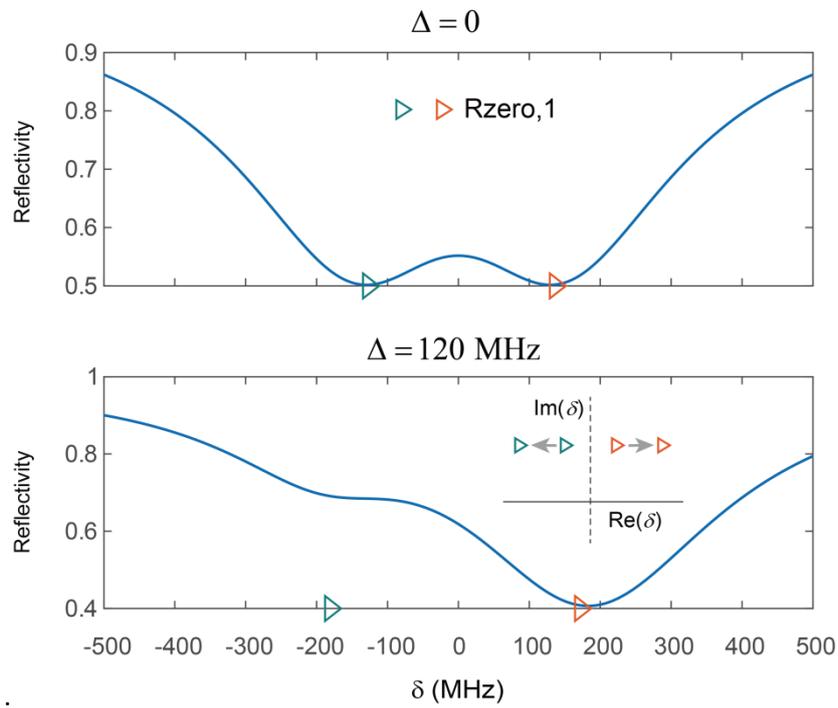

Fig. S3. Reflection spectra with/without detuning when $\gamma_1 = \gamma_{c1} + \Gamma_2$. The R-zeros move horizontally but the two reflection peak becomes asymmetric.

## S5. RSM existence conditions with cavity detuning

As dictated in the main text, the $RSM_{1,2}$ exist when $\kappa \geq \kappa_{1,2}$ for the detuned coupled models. Here we analyze the number of RSMs with different parameters. RSMs call for purely real $\delta_{\text{R-zero}}$ (Eq. (S12)), it requires that:

$$\Delta^2 = \frac{(\gamma_1 - \gamma_{c1} + \Gamma_2)^2(\kappa^2 - \kappa_1^2)}{4\Gamma_2(\gamma_{c1} - \gamma_1)} \tag{S13}$$

Where $\kappa_1 = \sqrt{(\gamma_{c1} - \gamma_1)\Gamma_2}/2$ is the RSM condition for the port 1 without detuning. Eq. (S10) ensures that the RSMs with detuning demands:

$$\begin{aligned}\gamma_{c1} &> \gamma_1 \\ \kappa &\geq \kappa_1\end{aligned} \tag{S14}$$

The first condition of Eq.(S14) is the overcoupling requirement inherited from the no detuning situation. The second condition indicates that when $\kappa$ exceeds the criterion of the RSM without detuning, it is possible to achieve one or more RSMs by inducing detuning at (as illustrated in Fig. 2(d) of the main text and Fig. S4):

$$\delta_{\text{RSM}} = \frac{2\Delta(\kappa)(\Gamma_2 + \gamma_{c1} - \gamma_1)}{\Gamma_2 - (\gamma_{c1} - \gamma_1)} \tag{S15}$$

Conversely, for $\kappa < \kappa_1$, RSM is no longer achievable. This result indicates that there is only one RSM for a given $\Delta$ except for the critical coupling scenario.

Fig. S4 shows the characteristics of the reflection spectrum of the port 1 with cavity detuning but ignore the incident cavity loss ($\gamma_1 = 0$). The RSM numbers under different parameters has analyzed by the R-zero plots in the main text, here exhibit additional properties of RSMs. On the one hand, it is clearly that the RSM existence condition requires $\kappa \geq \kappa_1$ otherwise there could only exist one R-zero at $\Delta = 0$ and $\delta = 0$. On the other hand, when total loss ($\Gamma = \gamma_{c1} + \Gamma_2$) holds and $\gamma_{c1} = 0.5\Gamma$ (critical coupling), the reflection map exhibits a symmetric behavior against $\Delta$ and the RSM always occurs at $\Delta = 0$.

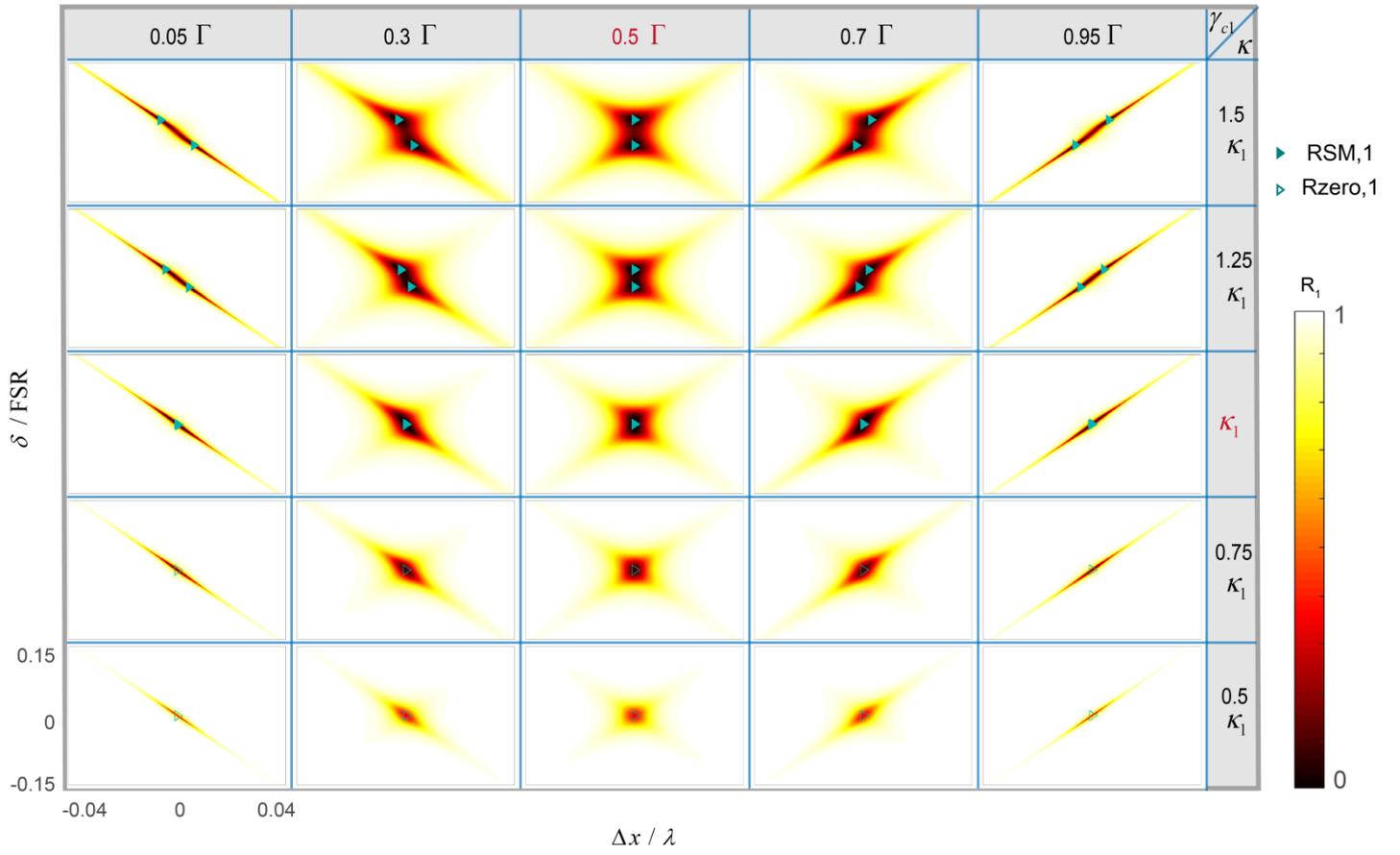

Fig. S4. Reflection map of the port 1 with detuning, as a function of the external coupling and inner coupling. $\gamma_{c1} = 0.5\Gamma$ and $\kappa = \kappa_1$ are obviously the phase transition boundaries for RSMs. The detuning is indicated by $\Delta x$. In particular, the square reflection spectrum located in the center arises when $\gamma_1 = 0$.